# Housing Search in the Age of Big Data: Smarter Cities or the Same Old Blind Spots?

Geoff Boeing[1], Max Besbris[2], Ariela Schachter[3], John Kuk[3]

**Abstract:** Housing scholars stress the importance of the information environment in shaping housing search behavior and outcomes. Rental listings have increasingly moved online over the past two decades and, in turn, online platforms like Craigslist are now central to the search process. Do these technology platforms serve as information equalizers or do they reflect traditional information inequalities that correlate with neighborhood sociodemographics? We synthesize and extend analyses of millions of US Craigslist rental listings and find they supply significantly different volumes, quality, and types of information in different communities. Technology platforms have the potential to broaden, diversify, and equalize housing search information, but they rely on landlord behavior and, in turn, likely will not reach this potential without a significant redesign or policy intervention. Smart cities advocates hoping to build better cities through technology must critically interrogate technology platforms and big data for systematic biases.

## 1. Introduction

Housing search technologies are changing and, as a result, so are housing search behaviors. The most recent American Housing Survey revealed that, for the first time, more urban renters found their current homes through online technology platforms than any other information channel. These technology platforms are online services cast in the role of information distributor, collecting and disseminating user-generated content and constructing a virtual agora for users to share information with one another. Because they can provide real-time data about various urban phenomena, these platforms are a key component of the smart cities paradigm (Batty 2012;

Crittenden 2017; Lim, Kim, and Maglio 2018; Shaw 2018; van der Graaf & Ballon 2018).

This paradigm promotes technology platforms as both a technocratic mode of monitoring cities and a utopian mode of improving urban life through big data (Batty 2012). In this context, big data typically refers to massive streams of user-generated content resulting from millions or billions of decentralized human actions. Data exhaust from Craigslist and other housing technology platforms offers a good example: optimistically, large corpora of rental listings could provide housing researchers and practitioners with actionable insights for policymaking while also equalizing access to information for otherwise disadvantaged homeseekers (Boeing and Waddell 2017; Boeing et al. 2020). But how good are these platforms at resolving the types of problems that already plague old-fashioned, non-big data? Does this broadcasting of information reduce longstanding geographic and demographic inequalities or do established patterns of segmentation and sorting remain (Ellen et al. 2016; Faber 2018; Jargowsky 2018; Peterson and Krivo 2010; Sampson 2012)?

This article draws together two ongoing research projects investigating online rental listings whose results, when put into conversation with one another, shed new light on this housing-technology context. It extends them with a new empirical component and theorizes the findings in light of the smart cities paradigm. Through a mix of computational and statistical analyses of millions of Craigslist rental listings, we find substantial variation in terms of different volumes and types of information supplied in online listings, correlated with neighborhood demographics. Majority-white neighborhoods are over-represented online while poor and minority neighborhoods have disproportionately fewer listings. Listings in higher-poverty neighborhoods contain less information compared to listings in lower-poverty neighborhoods, and advertisements in predominantly Black or Latino neighborhoods—regardless of poverty levels—contain less information about housing units and neighborhood amenities than listings in neighborhoods with higher percentages of white residents.

In light of recent scholarship stressing access to information as key to understanding disparate behaviors and outcomes in the housing search (Carrillo et al. 2016; Krysan and Crowder 2017; Marr 2005; Rosen 2014; Rosenblatt and DeLuca 2012), our work suggests that policies to reduce spatial inequality—particularly as it results from residential segregation—should focus on *how* individuals find homes and make attempts to equalize the information available to homeseekers. Additionally, we argue that in their current form online platforms may reproduce and even intensify existing forms of inequality within cities (Angelo and Vormann 2018; Brannon 2017). As long as technologies rely on user-generated content (such as advertisements composed by landlords) and user behavior (such as landlords' decisions about whether and how to list units) they are, alone, unlikely to promote social equity for citizens or produce representative datasets for policymakers. They also present novel policy problems that cannot be addressed with conventional housing anti-discrimination laws. If technology advocates aim to reduce housing inequalities, they must recognize the realities of the search process, the information pipeline, and both the potentials and limitations of online platforms in constructing equitable—and smart—cities and producing evenly-representative knowledge for policymaking.

## 2. Housing Search Technologies and Filter Bubbles

### 2.1. Housing Search

Understanding how individuals choose homes is key to understanding sociospatial processes like neighborhood change and residential segregation. Dominant approaches to explaining residential outcomes stress a mix of factors—including individuals' economic capacities, their preferences, and discrimination—in determining where they live (Crowder and Krysan 2016). Not all homeseekers can afford to live in all places, they may have preferences for certain amenities, geographies, or neighborhood demographics, and some housing purveyors discriminate in ways that drive homeseekers to locate in one neighborhood over another. Recent work, however, has stressed that homeseeking occurs in complex information environments that condition how individuals search for homes and where they ultimately live.

In their social structural sorting perspective, Krysan and Crowder (2017) draw on advances in decision-making science to argue that the housing search is multi-staged and iterative. More broadly, they show how existing racial/ethnic residential segregation produces further segregation. Before they begin a search, individuals have a circumscribed set of places they know about. Since social life is highly segregated by race/ethnicity (McPherson, Smith-Lovin, and Cook 2001), this knowledge, gleaned from friends, coworkers, family, media, and social experiences, is usually limited to places where the majority racial/ethnic group matches the searcher's. Residents know little about places other than the one in which they currently live, and they tend to know more about neighborhoods where demographics largely match those of their current place of residence and where their particular racial/ethnic group is the majority (Krysan and Bader 2009).

While past work has shown that the search strategies of different racial/ethnic groups vary (Farley 1996; Krysan 2008), all homeseekers have their choice sets filtered by their social networks before they start comparing units and are highly influenced by advice from peers (Lareau 2014). When homeseekers begin the search, they further limit their choice set by avoiding neighborhoods where they think they might face discrimination. Housing market intermediaries like real estate agents, mortgage brokers, community groups, and landlords also serve to limit the kinds of places searchers might consider and often use race/ethnicity and other characteristics like class, occupation, cultural tastes, or housing voucher status to further narrow options (Besbris 2016; Besbris and Faber 2017; Krysan and Lewis 2014; Korver-Genn 2018; Rosen 2014; Roscigno, Karafin, and Tester 2009; Walter and Wang 2016).

Key to this understanding of the housing search is that the search tools themselves are highly determinate: homeseeking tools, behaviors, and outcomes are all linked. Search tools include general information gathered from social networks and past experiences, but also particular media. The internet has transformed the housing search process, introducing various websites that provide information on available housing (Rae 2015). In the context of contemporary theories of housing search, this wider availability of data could improve overall access to information and interrupt the reproduction of existing forms of segregation by providing searchers with information on places they might not otherwise consider or even know about.

However, little research to date examines the actual information on these sites. It may be the case that new information technologies designed, in part, to improve aspects of urban life (like the housing search) in fact reflect or even exacerbate existing inequalities. In other words, if online platforms provide information that varies along existing dimensions of inequality, they could intensify segregation by in effect steering particular demographic groups toward or away from particular places.

## 2.2. Technology Platforms and Filter Bubbles

While Craigslist and other housing information sharing platforms have quickly claimed a central role in rental markets, our empirical knowledge about them has lagged behind (Schachter and Besbris 2017; Boeing and Waddell 2017). Little research has examined the actual supply of information on available housing. While we know a great deal about how homeseekers narrow their searches in ways that reproduce existing patterns of segregation and difference, we know far less about what types of information are available to them.

Online housing platforms could hypothetically alleviate traditional barriers to accessing information on available housing units (Steil and Jordan 2017). In particular, rental listing websites have the potential to reduce search costs while broadening the number and types of neighborhoods and units that housing seekers can explore (McLaughlin and Young 2017). For example, Craigslist, the most-used rental listings website, allows landlords to create listings with the location, rent, and amenities of a particular unit. Any searcher can access these listings: there is no cost for posting or searching. Free, publicly-available platforms like Craigslist could help introduce seekers to units in neighborhoods with different demographic profiles than those of the searcher, overcoming traditional social network and structural information filters that shape and segregate the search process (DeLuca and Rosenbaum 2003; Farley 1996;

Krysan 2008; Krysan and Crowder 2017; Rosen 2014; Rosenbaum, Reynolds, and DeLuca 2002; Schwartz et al. 2017).

However, if the information available online is unequal (e.g., housing seekers in whiter and wealthier neighborhoods have access to a surplus of useful information online while seekers elsewhere continue to face a relative deficit) then housing technology platforms alone will not reduce inequality in the housing market. Different communities may rely on different information channels for various reasons, but as housing information supplies continue to move online, self-selection into online information sharing platforms could replicate structural sorting mechanisms underlying residential segregation. Without exposure to a broad set of diverse neighborhoods, searchers might enter "filter bubbles" (Flaxman, Goel, and Rao 2016). This term describes the information isolation and segregation that can form when an individual's technology-mediated search habits filter information, tailored to the searcher, that ostensibly provides a holistic window into the real world but actually suffers from significant biases. This, in turn, circumscribes individuals' knowledge and expectations of reality while constructing the false impression of a comprehensive understanding. Information technologies can narrow knowledge and shift public opinion in ways that otherwise would not occur if individuals were exposed to the broader information being filtered out (O'Neil 2016).

Filter bubbles have emerged as a concern across the social sciences (Bakshy, Messing, and Adamic 2015; Garrett 2009). In online housing platforms, information filtering would act like other steering mechanisms—providing information unequally and selectively—leading to different search behaviors and outcomes for different types of homeseekers. Searchers' exposure to systematically different, place-based information would shape their understandings of the market and of particular neighborhoods. Some neighborhoods would have their reputations as places with a range of housing options and amiable landlords reified, while others—where less or even hostile information is provided in listings—would come to be known as inhospitable with units of lower quality.

Varying quality of information supplied online could particularly impact vulnerable communities. "Websites with low-income housing listings lack information on neighborhood characteristics" (Bergman 2018)—meaning that any searcher, advantaged or disadvantaged, looking at ads for housing in a disadvantaged neighborhood may have less information on which to base their housing decisions. The lack of information in certain neighborhoods may drive some renters away as homeseekers may bypass listings with less information. If these listings are concentrated in disadvantaged neighborhoods then the particular amenities of disadvantaged places will be hidden from searchers, and those who can afford to may look elsewhere for housing. If differences in information track with other neighborhood-level demographic characteristics, like race/ethnicity or socioeconomic status, that already evoke positive or negative assessments of neighborhoods (Besbris et al. 2015; Quillian and Pager 2001; Sampson and Raudenbush 2004), then online housing platforms will further filter homeseekers' knowledge and steer them toward or away from certain places. They would fail to live up to the promise of decentralized technology platforms for building smarter, more-equitable cities.

## 3. Assessing Online Platforms' Information Supply

To date, little research has analyzed Craigslist's representativeness or sociospatial biases. On one hand, as a publicly available and free technology platform, Craigslist offers exceptionally low barriers to entry. Unlike newspaper listings or brokers, it requires no payment from landlords to list their own units. With such low barriers to entry, Craigslist could possibly be the most representative exchange of rental information, although use is contingent on access and ability to navigate the internet. While alternative information channels exist for luxury rental listings (various local websites and specialized brokers), low-income listings (including gosection8.com), and non-English listings (including language-based groups on Facebook and websites like apartamentos.com), Craigslist is by far the most trafficked and largest single source of rental information.

To assess its equalizing potential, we review, empirically extend, and retheorize recent/ongoing research projects that collected Craigslist rental listings via web scraping. First, we review our recent assessment of over/under-representation; second, we review our recent assessment of information quantity/quality; and third, we conduct a new empirical analysis of information provision relevant to filtering.

### 3.1. Listing Over- and Under-Representation

Boeing & Waddell (2017) collected every rental listing posted in every Craigslist subdomain across the US between May and July 2014, then removed duplicates and listings without a geocode to produce a final clean dataset of 1.4 million georeferenced listings (full methodological details in Boeing & Waddell 2017). Craigslist listings contain information about advertised rent and optional details—when provided by the lister—including the number of bedrooms, square footage, and descriptions of the surrounding neighborhood. Listers include professional landlords, individual homeowners, agents, and brokers, but Craigslist provides no information about these listers or their demographics. Unit geolocation is present when listers provide it by dropping a pin on a web map. This avoids many of the challenges inherent in address geocoding by capturing the lister's stated location of the unit for rent.

Boeing (2019) assesses over- and under-representation on Craigslist in the 12,505 census tracts in the core cities of the US's 50 largest metropolitan areas by comparing average volumes of listings per tract to the average volume of vacant units for rent, per the American Community Survey, then estimates spatial regression models including MSA fixed effects to investigate *ceteris paribus* relationships between Craigslist representation and sociodemographic and built environment predictors. Finally, Boeing conducts tests on these variables between over- and under-represented tracts to identify demographic differences in neighborhoods that Craigslist relatively over- or under-supplies information on.

This analysis reveals that Craigslist listings concentrate in whiter, wealthier communities. Over half of majority-White tracts are over-represented on Craigslist (relative to the expected listing volume based on vacancy rates) while less than a quarter of majority-Black or Latino tracts are. Over-represented tracts have a White proportion of the population 20 percentage-points higher, on average, than under-represented tracts. Meanwhile, the Black proportion is 14 percentage-points lower and the Latino proportion is 8 percentage-points lower. In over-represented tracts, average incomes are $21,000 higher and average home values are $80,000 higher, while the proportion of the population with a bachelor's degree or higher is 17 percentage-points higher. Controlling for covariates, higher incomes and education levels predict greater representation on Craigslist, while greater Black or Latino proportions of the population predict lower representation, all else equal. In sum, these findings suggest that this technology platform offers larger choice sets and reduced search costs for homeseekers in whiter, wealthier communities, drawing into question its efficacy as an information-equalizing civic technology.

### 3.2. Per-Listing Information Quantity/Quality

Besbris et al. (2018) collected all listings posted in the Craigslist subdomains of the 50 largest US metropolitan areas on a weekly basis between May 2017 and February 2018, then cleaned them to remove duplicates and obviously invalid listings and to retain only geolocated listings, resulting in a final data set of 1.7 million listings. They test for differences in the amount and type of information contained within listings. They first examine the number of optional information categories provided, the number of images included, and the overall number of words in each listing to capture differences in information available to searchers in census tracts with varying racial/ethnic and poverty compositions (see Wang et al. 2018). They then estimate differences in each of these measures across a sociodemographic typology of tracts, then use computational text analysis techniques, including structural topic modeling, to examine differences in the kinds of information included in tracts' listings. Structural topic models uncover the type and the prevalence of topics (i.e., collections of words that share a common theme) in each listing. Through these models, they discover prevalent themes exist in the text and estimate which themes are likely to be found in certain neighborhoods.

Besbris et al. find that listings in communities with more Black, Latino, or poorer residents contain less information. Further, the information in listings in tracts with more Black, Latino, or poorer residents disproportionately focuses on tenant (dis)qualifications (e.g., proof of income, eviction history, criminal history) rather than unit/amenity descriptions. In contrast, listings in whiter or lower-poverty areas contain more information and devote more text to describing units/amenities. A clear relationship exists between tract poverty and the amount and type of information provided, but there is also a racial hierarchy: listings in low-poverty Black or Latino communities contain less information and a stronger focus on tenant (dis)qualifications compared to similarly low-poverty White communities. Listings in White tracts have a more extensive discussion on neighborhood amenities (e.g., proximity to parks and restaurants, public transportation availability) than Black or Latino counterparts.

### 3.3. Word Count by Race and Poverty Status

To further understand where searchers are likely to find more or less information through online listings, we extend this previous analysis here in Tables 1 and 2. Table 1 reports average descriptive differences in word count among listings, by tract race and poverty status. For example, the first row shows that listings in poor White, Black, and Latino tracts, as well as non-poor Black and Latino tracts, contain fewer words on average and thus provide less overall information than listings in non-poor White tracts. Listings in poor Black tracts contain fewer average words than any other tract type, and even non-poor Black tracts contain fewer than all other tract types except poor Black and Latino tracts. In contrast, listings in poor and non-poor Asian tracts contain more words on average than those posted in all other tracts, including non-poor White tracts. Table 1 depicts information disparities that cannot be explained by tract poverty status alone. Rather, per-listing information volume varies at the intersection of race and poverty, leaving poor Latino and, particularly, Black tracts the most relatively-disadvantaged in terms of information content, and non-poor Asian tracts the most advantaged.

The differences by tract race/ethnicity and poverty described above are measured at the aggregate level across all tracts in the 50 largest MSAs and may therefore mask heterogeneity across metropolitan markets. For example, information disparities may be associated with segregation rates and/or the relative supply of and demand for rental housing within metropolitan areas, which could influence how much and what types of advertising content are deemed necessary by landlords. In addition, variation in regional or municipal ordinances regulating rental housing, such as Seattle's recently struck-down law that required landlords to clearly state all tenant requirements in their advertisements, may contribute to information differences or the lack thereof. To explore these possibilities, Table 2 presents the Craigslist markets with the largest and smallest gaps in per-listing word count between White and Black non-poor tracts. Some markets, including Salt Lake City and Riverside, exhibit much larger White-Black gaps than the overall average of +32 words. Others, like Pittsburgh and Tampa, exhibit virtually no gap. Finally, while nationwide listings in non-poor Black tracts contain fewer words per listing on average than non-poor White tracts, we find a few metropolitan areas, such as Seattle, where this pattern reverses. While fully exploring the causes of these differences is beyond the scope of this paper, the patterns highlight the importance of examining different types of metropolitan areas to identify potential policy levers that may ameliorate (or exacerbate) inequality in online information exchanges. In total, these findings suggest how user-generated big data sets do not necessarily equalize information provided across geographies.

### 3.4. Presence of Filterable Fields

While this expands our understanding of sociospatial inequality online, some questions remain. Craigslist requires listers to provide a minimal amount of information, including a title, zip code, descriptive body (as much or as little text as the lister wants), and rent. Beyond these requirements, Craigslist allows listers to provide substantially more information via write-in options and checkboxes that homeseekers with different needs can use to filter listings. What is the relationship between neighborhood demographics and the distribution of this additional information? The presence or absence of these optional information fields shapes individual searchers' perspectives on the local rental market. If information is key to the search process, then any differences could impact where individuals are more or less willing to search. Less information in certain kinds of neighborhoods could drive away homeseekers who might otherwise be willing to consider living there.

To begin answering these questions we conduct a new analysis of additional information fields in listings that can be used to filter search results. We focus on three optional information fields essential to many searchers:

1) Whether an exact *address* is given for the listing. This is an optional field the landlord writes-in, rather than a set of check boxes. Compared to simply checking a box, providing this information requires slightly more effort by the landlord, signaling the landlord's willingness to communicate and level of trust (by providing an exact address rather than only a more general location).

2) Whether the listing indicates that a *washer/dryer* is available or not. This information field uses a set of check boxes that include an option to say "no laundry." Thus, while this information's (affirmative or not) presence may correlate with amenity presence, it offers multiple options, such that regardless of amenity presence all landlords could provide this optional information. However, checking a box about laundry requires less effort than writing in the exact address.

3) Whether the listing indicates that it *allows pets*. Unlike the *washer/dryer* field, there is no check box option to say "no pets", so this field might look quite different from the *washer/dryer* or *address* information fields, as its presence may depend on underlying amenities.

We select these three information fields because of their different characteristics, to explore information-provision inequality rather than underlying heterogeneity in the presence of amenities themselves. The resulting indicators denote the presence or absence of each information field. Importantly, they thus derive from information fields organized by Craigslist rather than unstructured text contained in the listing's body. Because these are categorical indicators, searchers have the option to filter their results using each of these categories. In other words, these fields are both substantively important to searchers (e.g., matching the needs of pet owners) and they filter which listings searchers even see if their housing needs include such requirements. While all listings have an address and landlord unwillingness to provide it may communicate some apprehension about engagement with potential tenants, the other two outcomes are more complicated. Indeed, the availability of laundry or accommodation of pets may correlate with location because certain types of housing stock allow for them. Housing quality, as opposed to neighborhood demographics, may be more determinate for these outcomes. Nevertheless, because landlords have the option of checking a box that indicates laundry is not available (for instance), not providing *any* information either way could affect homeseekers' understanding of the legibility and quality of different neighborhoods' housing stock and landlords.

To test for tract-level sociodemographic differences in the provision of these optional information fields, we estimate three Linear Probability Models of the probability that a given listing contains each field. We use Linear Probability Models instead of logit or probit models because our estimation includes fixed effects and our quantity of interest is differences in the three optional informational fields across tracts, not predicted probabilities. We compute cluster-robust standard errors to address heteroskedasticity concerns. As in Besbris et al. (2018), we operationalize an 8-category sociodemographic typology of tracts. Each model contains MSA fixed effects and controls for unit rent, tract percentage of population with a bachelor's degree, percentage of units that are occupied by renters, percentage of units built after 2014, percentage of units that are vacant, and foreign-born percentage of population.

Table 3 shows that, relative to listings in non-poor White tracts, listings in poor Black and Latino tracts are significantly less likely to include an exact address, indicate washer/dryer availability, or indicate if they allow pets. Compared to non-poor White tracts, listings in poor White tracts are about 4 percentage points less likely to contain an exact address. This difference is greater for listings in poor Black (7 percentage points less likely), Latino (9 percentage points less likely), and Asian (7 percentage points less likely) tracts. Interestingly, controlling additionally for percent foreign-born increases the gap between non-poor White and poor Latino tracts, a finding which deserves future research. We see similar trends for information on washer/dryer availability and allowing pets, though the effect sizes vary and a few are not statistically significant. However, it is not merely listings in poor tracts (where we might expect fewer amenities like laundry to be available) that are less likely to contain these information fields. Instead, we again find a clear racial hierarchy: relative to non-poor White tracts, non-poor Black tracts are less likely to provide information on washer/dryer availability or whether pets are allowed, and non-poor Latino tracts are less likely to provide an exact address or information on whether pets are allowed.

Although these three information fields have different characteristics, we discover similar patterns across their provision. Thus, these trends suggest a more nuanced form of information inequality rather than simply underlying differences in amenity presence. These fields contain information essential to the searches of many types of renters. Their disproportionate absence in Black and Latino neighborhoods could discourage White renters from considering suitable units in these communities, perpetuating residential segregation along racial lines through information inequality.

# 4. Smarter Cities or the Same Old Blind Spots?

## 4.1. Information Segregation and Filters

Data biases both influence policymakers' knowledge and create information inequalities that impact housing searches. In tandem, our analyses elucidate the disparate quantity and quality of information and potential filtering mechanisms in online rental listings, revealing important challenges for policymakers looking to use data generated from these platforms and painting a unified portrait of unequal access to information online.

We find significant sociospatial differences in both the volume and type of information provided to prospective tenants on Craigslist. Recent work has speculated about the potential of technology platforms to democratize information and broaden homeseeker choice sets, but our findings question these platforms' ability to make searches more equitable: online rental listings reproduce historical patterns of residential steering, sorting, and (information channel) segregation as a function of existing population distribution and inequality. Given the segregated nature of the housing search process (Krysan and Bader 2009; Krysan and Crowder 2017), our findings demonstrate that online housing listings are more likely to exacerbate rather than ameliorate inequality.

For example, if Black homeseekers are more likely to search for online listings in neighborhoods with more Black residents, they will on average view fewer available units and the listings that they view will include less information about living conditions and lease terms compared to a homeseeker searching in a White neighborhood. Alternatively, searchers who require particular amenities will be systematically unable to find essential information in poorer neighborhoods and in neighborhoods with higher shares of Blacks or Latinos, possibly preventing them from moving to these places. Indeed, decision making in various markets, including the housing market, occurs in stages where choices are filtered out quickly (Krysan and Crowder 2017; see also Bruch and Feinberg 2017). When renters searching for a unit with a washer and dryer use the available filtering tools on Craigslist, the platform will return few options in disadvantaged neighborhoods. In other words, Craigslist quickly forecloses searches in disadvantaged neighborhoods for homeseekers in need of particular amenities and with the means to search in more-advantaged neighborhoods. These searchers will fail to learn about disadvantaged places, and the knowledge they rely on for future moves and pass onto their social networks will remain siloed. In short, our evidence suggests that online search tools are part of, not a solution to, the segregated and iterative ways homeseekers experience housing searches.

The original goal of Craigslist was to create a free, equal platform open to all, but in practice we find uneven representation correlated with race/ethnicity and poverty. Indeed, we focused on Craigslist here because compared to other housing technology platforms it is the largest and most democratic in that it has minimal barriers to entry and no listing costs. Yet despite being the most accessible platform— and despite having no complicated algorithms that target search results—we still find that user-generated information reproduces traditional information segregation patterns. Other online housing platforms potentially distribute information in even more unequal ways, constructing filter bubbles in the residential search and sorting process. In fact, the US Department of Housing and Urban Development (HUD) recently sued Facebook for violating fair housing laws, claiming their platform limits who can see advertisements for housing based on their race/ethnicity, religion, and current location (Benner, Thrush, and Isaac 2019; Porter et al. 2019). As cities and citizens increasingly turn to technology platforms to mediate urban processes, more unanticipated consequences, such as these housing information filter bubbles, will likely appear.

However, regulating such inequalities will not be easy. The information inequalities we document here are particularly pernicious because they result from aggregate individual decisions that do not violate current fair housing laws that regulate individuals' words and behavior: it is not illegal for landlords to selectively share information about their units or neighborhood amenities on search platforms. Disparate impact laws could potentially be used to regulate filters shown to steer particular groups, but a recently proposed HUD rule seeks to raise the bar of proof for housing discrimination, which would further limit policymakers' ability to address information inequality (Badger 2019).

## 4.2. Smart Cities: Power and Blind Spots

New urban technology platforms and their data both reflect our world and shape it. Despite the optimistic rhetoric of smart cities advocates, techno-utopian solutions are not always equalizing, in practice or even in design. Cities increasingly rely on technology platforms as a mode of governance, administration, observation, and participation—but, in many ways, this reliance can reproduce or even exacerbate preexisting inequalities. While these platforms reflect millions of disaggregate transactions among rental market participants, they also (re-)construct the market itself. Online listings can reduce housing search costs, expand search radii without requiring physical location visits, and broaden homeseeker choice sets. But we find that housing information quantity and quality vary between neighborhoods, correlated with their sociodemographic profiles. In turn, the information-broadcasting benefits of these housing technology platforms are unevenly distributed among these communities.

On one hand, this unevenness concentrates the technology's benefits in privileged communities. On the other hand, it could also open up such communities to information-deprived housing seekers by making whiter, wealthier, and better-educated communities more-equally legible to everyone in the search process through a larger volume of rental listings and higher-quality unit information. But what opportunities for lower socioeconomic status families are opened up if these desirable communities' listings remain expensive or are perceived as unwelcoming to poorer or non-White homeseekers? In addition, substantial variation exists in the information inequality between cities, suggesting that these platforms serve homeseekers in different ways and might also serve policymakers (looking to harvest timely rental market information) in different ways.

Given the theoretical potential of housing technology platforms to broaden and diversify information for seekers, what can policymakers do to advance more-equal benefits while limiting the drawbacks of information segregation, filtering, and self-selection? Cities and tech companies have several possible avenues to consider jointly pursuing. For example, cities could legally require landlords to fill out more information fields when listing online. Voucher program staff could work with voucher holders on devising successful online search strategies, including encouraging searching by unit-level information (to break out of geographic silos) and providing more information on neighborhoods previously unknown to the searcher.

But focusing solely on search behavior may not be enough. Beyond requiring more information from listers on existing platforms, cities might require all landlords to register their leases with the municipality to create their own centralized housing website. Such websites could standardize information across listings and be designed to foster searching across wider sets of neighborhoods. Some cities already maintain databases of all properties leased or for-lease and this information could be made searchable. City/state affordable housing platforms like San Francisco's DAHLIA, Minnesota's HousingLink, and Massachusetts' (in-development) Housing Navigator demonstrate possibilities for connecting disadvantaged homeseekers to useful housing information (n.b. HousingLink collects Craigslist listings by hand to add to its own platform). However, dedicated information channels for different searcher groups may further entrench the information segregation we found on Craigslist. But as our findings reveal, information is not equally distributed across places, so any efforts to broaden knowledge about different neighborhoods would be welcome.

## 5. Conclusion

This article drew together research investigating Craigslist's online rental housing market and, using new analyses, theorized the impacts of technology platform-mediated housing search. It described the uneven quantity and quality of online information correlated with neighborhood demographics, creating unequal housing information supplies based on where you search. This calls into question the ability of technology platforms to serve as utopian, democratizing, equalizing forces when they rely on human content creation and preexisting sociospatial relations. We agree with Krysan and Crowder's (2017) recent call for broadening housing search information sources to lessen traditional sorting mechanisms. However, we remain skeptical of current technology platforms' ability to accomplish this goal without a significant redesign—since they reproduce traditional human tendencies—or even a restructuring of the capitalist logics that underpin them.

　　In the smart cities mode of urban monitoring, data exhaust can provide a window into what/how landlords are listing, and our findings point to important future research in this space. First, researchers should explore patterns on other housing platforms. Second, future work should employ mixed methods to investigate landlord intent and how homeseekers perceive listing information. How are prospective renters interacting with and receiving information from the platforms? What kinds of applicants do landlords expect and desire when they create their listings? Local market conditions likely play a role in how landlords create listings and what types of information they provide. Higher demand may reduce the pressure to compose long and detailed listings. Finally, we found that not all metropolitan areas are "equally unequal." Future research should unpack which aspects of these metropolitan areas explain their more or less equal information provision across neighborhood types. But by descriptively documenting the *existence* of such heterogeneity, we hope to motivate researchers and policymakers to investigate these metropolitan areas to ascertain why so much information inequality exists in some markets and so little in others. Comparisons of these places may yield valuable insights into potential policy solutions.

Overall, the differences documented here quantify how online platforms with user generated content do not automatically smooth information exchange, reduce information asymmetries, or attenuate entrenched sociospatial inequalities. Craigslist data do not capture all neighborhoods—nor the experiences of all renters—equally well. Importantly, Craigslist data alone do not allow policymakers to fully understand the experiences of renters searching for housing in lower income or minority neighborhoods. This is not simply a data problem. Although Craigslist's biases may be unintended, they nevertheless point to the broader limits of housing technology platforms themselves—which rely on user self-selection and structural market forces to generate information.

Policymakers focused on using technology platforms to reduce inequality and serve lower income and minority renters might operationalize Craigslist data to understand information gaps—including unit location, layout, and square footage—and biases experienced by disadvantaged housing seekers through both explicit and implicit forms of discrimination. Policymakers could even use insights from online listings data to work with technology firms to regulate, redesign, or propose alternative housing platforms that better serve low income and minority communities to share the internet's search cost reductions and information-broadcasting benefits more evenly among all homeseekers. But as our findings show, policymakers and technologists cannot rely on user-generated big data alone to meet all citizens' needs. Data sources must be continually analyzed and interrogated critically, as well as regularly compared for quality and breadth.

## Acknowledgements

The authors wish to thank Paul Waddell, who graciously allowed the reuse of data originally collected in his lab at UC Berkeley for some of these analyses.

**Table 1.** Pairwise differences in number of words per listing, across tract types

|  | White Non-poor | White Poor | Black Non-poor | Black Poor | Latino Non-poor | Latino Poor | Asian Non-poor | Asian Poor |
|---|---|---|---|---|---|---|---|---|
| White Non-poor | - | -22.1 | -32.4 | -54.9 | -9.9 | -33.0 | +24.3 | +10.4 |
| White Poor | +22.1 | - | -10.3 | -32.9 | +12.1 | -11.0 | +46.4 | +32.5 |
| Black Non-poor | +32.4 | +10.3 | - | -22.5 | +22.5 | -0.6 | +56.8 | +42.8 |
| Black Poor | +54.9 | +32.9 | +22.5 | - | +45.0 | +21.9 | +79.3 | +65.3 |
| Latino Non-poor | +9.9 | -12.1 | -22.5 | -45.0 | - | -23.1 | +34.3 | +20.3 |
| Latino Poor | +33.0 | +11.0 | +0.6 | -21.9 | +23.1 | - | +57.4 | +43.4 |
| Asian Non-poor | -24.3 | -46.4 | -56.8 | -79.3 | -34.3 | -57.4 | - | -13.9 |
| Asian Poor | -10.4 | -32.5 | -42.8 | -65.3 | -20.3 | -43.4 | +13.9 | - |

Note: Table values are computed by subtracting the average number of words used in the tract type in the row from the average number of words used in the tract type in the column. For example, the information in column 1, row 2 should be interpreted as listings in White non-poor tracts have 22.1 more words on average than White poor tracts. The tract racial group is based on its plurality racial group. We use a 30% poverty rate threshold to classify whether a tract is poor or non-poor. See Besbris et al. (2018) for more detail on the underlying data and methodology.

**Table 2.** Differences in number of words per listing between white non-poor tracts and black non-poor tracts, by MSA

| Top 5 MSAs | Difference |
|---|---|
| Salt Lake City | +177.04 |
| Riverside-San Bernardino-Ontario | +154.76 |
| San Diego-Carlsbad | +106.98 |
| Las Vegas-Henderson-Paradise | +85.96 |
| Providence-Warwick | +83.03 |
| **Most-equal MSAs** | |
| Pittsburgh | +3.17 |
| Tampa-St. Petersburg-Clearwater | +0.67 |
| Richmond | +0.13 |
| Kansas City | -0.54 |
| Miami-Fort Lauderdale-West Palm Beach | -2.91 |
| **Bottom 5 MSAs** | |
| St. Louis | -7.30 |
| New Orleans-Metairie | -9.32 |
| Sacramento-Roseville-Arden-Arcade | -9.37 |
| Cincinnati | -49.61 |
| Seattle-Tacoma-Bellevue | -63.14 |

Note: Table values are computed by subtracting the average number of words used in Black non-poor tracts from the average number of words used in White non-poor tracts for each MSA. For example, in Salt Lake City, listings in White non-poor tracts have 177 more words on average than listings in Black non-poor tracts. We classify a listing as belonging to a White non-poor tract when the listing is located in a tract that is plurality White and has a <30% poverty rate. We classify a listing as belonging to a Black non-poor tract when the listing is located in a tract that is plurality Black and has a <30% poverty rate. See Besbris et al. (2018) for more detail on the underlying data and methodology.

**Table 3.** Linear probability models predicting the existence of information fields

| | Dependent variable: | | |
|---|---|---|---|
| | Address Exists | Washer/Dryer | Allow Pets |
| | (1) | (2) | (3) |
| White Poor | -0.041* | -0.011 | -0.056** |
| | (0.017) | (0.016) | (0.017) |
| Black Non-poor | 0.017 | -0.033+ | -0.040* |
| | (0.015) | (0.018) | (0.016) |
| Black Poor | -0.068** | -0.041* | -0.056** |
| | (0.021) | (0.019) | (0.019) |
| Latino Non-poor | -0.042* | -0.018 | -0.055** |
| | (0.018) | (0.018) | (0.016) |
| Latino Poor | -0.086** | -0.051* | -0.080** |
| | (0.024) | (0.024) | (0.024) |
| Asian Non-poor | -0.017 | 0.023 | -0.022 |
| | (0.020) | (0.021) | (0.024) |
| Asian Poor | -0.071* | -0.051 | -0.111** |
| | (0.029) | (0.056) | (0.036) |
| Unit Price ($1,000) | 0.005* | 0.017** | 0.006** |
| | (0.002) | (0.002) | (0.001) |
| % College | 0.001* | 0.001** | 0.001** |
| | (0.0002) | (0.0002) | (0.0002) |
| % Units Renter | 0.002** | 0.0004* | 0.002** |
| | (0.0002) | (0.0002) | (0.0002) |
| % Built after 2014 | 0.001 | 0.001 | 0.005** |
| | (0.001) | (0.001) | (0.001) |
| % Foreign Born | 0.001** | -0.001* | -0.0005 |
| | (0.0004) | (0.0004) | (0.0004) |
| % Vacancy | -0.003** | -0.0001 | -0.003** |
| | (0.001) | (0.001) | (0.0005) |
| Observations | 1,694,297 | 1,694,297 | 1,694,297 |
| # of Census Tracts | 37,310 | 37,310 | 37,310 |
| $R^2$ | 0.087 | 0.080 | 0.050 |
| Adjusted $R^2$ | 0.087 | 0.080 | 0.050 |
| Residual Std. Error | 0.408 | 0.480 | 0.475 |

+$p<0.1$; *$p<0.05$; **$p<0.01$

Note: The dependent variables are dichotomous, indicating whether (1) optional exact address information is provided in the listing, (2) any check-box information exists regarding the availability of washer/dryer, and (3) the listing includes a check box allowing pets (either cats, dogs, or both). Tracts are classified according to the plurality racial group and a 30% poverty rate threshold to classify whether a tract is poor or non-poor. All 3 models include MSA fixed effects and standard errors are clustered at the tract level. Coefficients are relative to the white non-poor reference group. See Besbris et al. (2018) for more detail on the underlying data and methodology.